\documentclass{sigchi}

\CopyrightYear{2018}
\setcopyright{acmlicensed}
\conferenceinfo{CHI 2018,}{April 21--26, 2018, Montreal, QC, Canada}
\isbn{978-1-4503-5620-6/18/04}\acmPrice{\$15.00}
\doi{https://doi.org/10.1145/3173574.3173779}

\usepackage{balance}       
\usepackage{graphics}      
\usepackage[T1]{fontenc}   
\usepackage{txfonts}
\usepackage{mathptmx}
\usepackage[pdflang={en-US},pdftex]{hyperref}
\usepackage{color}
\usepackage{booktabs}
\usepackage{textcomp}

\usepackage{graphicx}
\usepackage{caption}
\usepackage{subcaption}
\usepackage{algorithm}
\usepackage{algpseudocode}

\usepackage{microtype}        
\usepackage{ccicons}          

\usepackage{todonotes}

\usepackage{enumitem}

\def\plaintitle{DeepWriting: \\ Making Digital Ink Editable via Deep Generative Modeling}

\def\emptyauthor{}
\def\plainkeywords{Handwriting; Digital Ink; Stylus-based Interfaces; Deep Learning; Recurrent Neural Networks}

\makeatletter
\def\url@leostyle{%
  \@ifundefined{selectfont}{
    \def\UrlFont{\sf}
  }{
    \def\UrlFont{\small\bf\ttfamily}
  }}
\makeatother
\def\pprw{8.5in}
\def\pprh{11in}

\definecolor{linkColor}{RGB}{6,125,233}
\hypersetup{%
  pdftitle={\plaintitle},
  pdfauthor={\emptyauthor},
  pdfkeywords={\plainkeywords},
  pdfdisplaydoctitle=true, 
  bookmarksnumbered,
  pdfstartview={FitH},
  colorlinks,
  citecolor=black,
  filecolor=black,
  linkcolor=black,
  urlcolor=linkColor,
  breaklinks=true,
  hypertexnames=false
}

\definecolor{caribbeangreen}{rgb}{0.0, 0.8, 0.6}
\definecolor{cambridgeblue}{rgb}{0.64, 0.76, 0.68}
\definecolor{celadon}{rgb}{0.67, 0.88, 0.69}
\definecolor{armygreen}{rgb}{0.29, 0.33, 0.13}
\definecolor{carminered}{rgb}{1.0, 0.0, 0.22}
\definecolor{americanrose}{rgb}{1.0, 0.01, 0.24}
\definecolor{amber(sae/ece)}{rgb}{1.0, 0.49, 0.0}
\definecolor{asparagus}{rgb}{0.53, 0.66, 0.42}
\definecolor{LightCyan}{rgb}{0.88,1,1}
\definecolor{LightGreen}{rgb}{0.77,0.93,0.8}
\definecolor{LightOrange}{rgb}{0.95,.67,0.47}
\definecolor{DarkYellow}{rgb}{0.478,.478,0.0058}

\newcommand{\modelname}{C-VRNN~}

\newif\ifshowchanges
\showchangesfalse
\ifshowchanges
    
\else
    
\fi

\usepackage{mathtools}

\DeclarePairedDelimiterX{\infdivx}[2]{(}{)}{%
  #1\;\delimsize\|\;#2%
}

\newcommand{\figref}[1]{Figure~\ref{#1}}
\newcommand{\refequ}[1] {Eq.~(\ref{#1})}
\newcommand{\refequs}[2] {Eq.~(\ref{#1}-\ref{#2})}

\definecolor{inputcolor}{RGB}{0,0,137}
\definecolor{inputlstmcolor}{RGB}{165,165,165}
\definecolor{latentlstmcolor}{RGB}{192,0,0}
\definecolor{outputlstmcolor}{RGB}{255,192,0}

\usepackage{amsmath}
\usepackage{amssymb}

\begin{document}

\title{\plaintitle}

\author{
	\fontseries{b}\fontsize{12}{13}\selectfont
	Emre Aksan \,
	Fabrizio Pece \,
	Otmar Hilliges 
	\\
	\fontseries{m}\fontsize{11}{12}\selectfont
	Advanced Interactive Technologies Lab, ETH Z\"urich \\
	\fontseries{m}\fontsize{11}{12}\selectfont
	\{eaksan, pecef, otmarh\}@inf.ethz.ch
}

\teaser{
	\centering{
		\includegraphics[width=\textwidth]{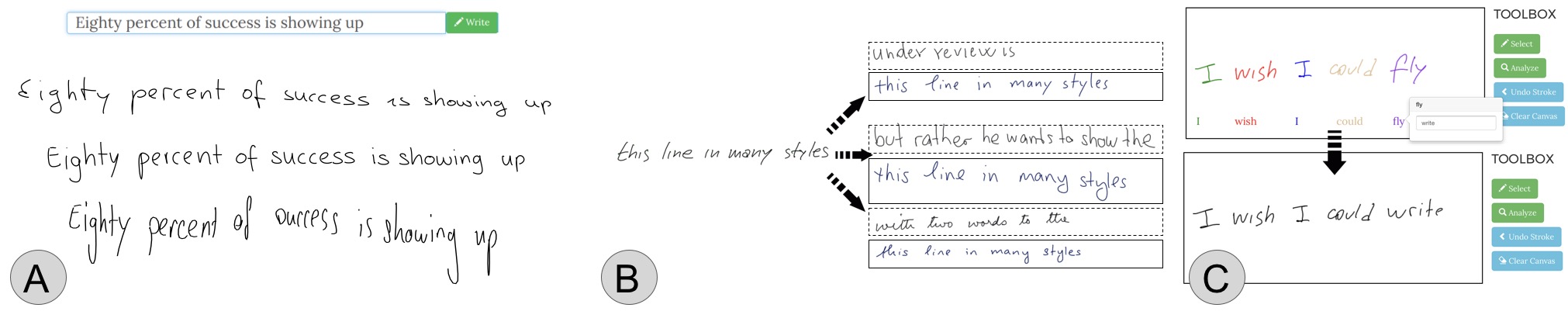}
		\caption{We propose a novel generative neural network architecture that is capable of disentangling style from content and thus makes digital ink editable. Our model can synthesize handwriting from typed text while giving users control over the visual appearance (A), transfer style across handwriting samples (B, solid line box synthesized stroke, dotted line box reference style), and even edit handwritten samples at the word level (C).}
		\label{fig:teaser}
		\vspace{-10pt}
	}
}

\maketitle


\begin{abstract}
Digital ink promises to combine the flexibility and aesthetics of handwriting and the ability to process, search and edit digital text. Character recognition converts handwritten text into a digital representation, albeit at the cost of losing  personalized appearance due to the technical difficulties of separating the interwoven components of content and style. In this paper, we propose a novel generative neural network architecture that is capable of disentangling style from content and thus making digital ink editable. Our model can synthesize arbitrary text, while giving users control over the visual appearance (style). For example, allowing for style transfer without changing the content, editing of digital ink at the word level and other application scenarios such as spell-checking and correction of handwritten text. We furthermore contribute a new dataset of handwritten text with fine-grained annotations at the character level and report results from an initial user evaluation.

\end{abstract}





\category{H.5.2}{User Interfaces}{Input Devices and Strategies}
\category{I.2.6}{Learning}{}

\keywords{\plainkeywords}


\section{Introduction}\label{sec:intro}
Handwritten text has served for centuries as our primary mean of communication and cornerstone of our education and culture, and often is considered a form of art~\cite{Robinson2007}. It has been shown to be beneficial in tasks such as note-taking~\cite{Mueller2014}, reading in conjunction with writing~\cite{Sellen2003} and may have a positive impact on short- and long-term memory \cite{Berninger2012}. However, despite progress in character recognition~\cite{Liwicki2007}, fully digital text remains easier to process, search and manipulate than handwritten text which has lead to a dominance of typed text.

In this paper we explore novel ways to combine the benefits of digital ink with the versatility and efficiency of digital text, by making it editable via disentanglement of style and content. Digital ink and stylus-based interfaces have been of continued interest to HCI research (e.g., \cite{Hammond2015, Hinckley2014, Riche2017, Xia2017, Zitnick2013}).
However, to process digital ink one has typically to resort to optical character recognition (OCR) techniques (e.g.,~\cite{EvernoteOCR2017}) thus invariably losing the personalized aspect of written text.
In contrast, our approach is capable of maintaining the author's original style, thus allowing for a seamless transition between handwritten and digital text.
Our approach is capable of synthesizing handwritten text, taking either a sequence of digital ink or ASCII characters as input.
This is a challenging problem: while each user has a unique handwriting style \cite{Srihari2002,Zhang2003}, the parameters that determine style are not well defined.
Moreover, handwriting style is not fixed but changes temporally based on context, writing speed and other factors~\cite{Srihari2002}.
Hence so far it has been elusive to algorithmically recreate style faithfully, while being able to control content. A comprehensive approach to handwriting synthesis must be able to maintain \emph{global} style while preserving \emph{local} variability and context (e.g., many users mix cursive and disconnected styles dynamically).

Embracing this challenge we contribute a novel generative deep neural network architecture for the conditional synthesis of digital ink. The model is capable of capturing and reproducing local variability of handwriting and can mimic different user styles with high-fidelity.
Importantly the model provides full control over the content of the synthetic sequences, enabling processing and editing of digital ink at the word level.
The main technical contribution stems from the ability to disentangle latent factors that influence visual appearance and content from each other.
This is achieved via an architecture that combines autoregressive models with a learned latent-space representation, modeling temporal and time-invariant categorical aspects (i.e., character information).

More precisely we propose an architecture comprising of a recurrent variational autoencoder \cite{Chung2015,Kingma2013} in combination with two latent distributions that allow for \emph{conditional} synthesis (i.e., provide control over style and content) alongside a novel training and sampling algorithm.
The system has been trained on segmented samples of handwritten text collected from 294 authors, and takes into account both temporal and stylistic aspects of handwriting. Further, it is capable of synthesizing novel sequences from typed-text, transferring styles from one user to another, editing digital ink at the word level and thus enables compelling application scenarios including spell-checking and auto-correction of digital ink.

We characterize the performance of the architecture via a thorough technical evaluation, and assess its efficacy and utility via a preliminary experimental evaluation of the interactive scenarios we implemented. Further, we contribute a new dataset that enhances the IAM On-Line Handwriting Database (IAM-OnDB) \cite{Liwicki2005} and includes handwritten text collected from 294 authors with character level annotations. Finally, we plan to release an open-source implementation of our model. 


\section{Related work}\label{sec:rw}
Our work touches upon various subjects including HCI (e.g., \cite{Hammond2015, Perteneder2015, Riche2017,Xia2017}), handwriting analysis~\cite{Pulver1972} and machine learning (e.g., \cite{Elarian2014, Graves2013, Kingma2013,Plamondon2000}).

\subsection{Understanding Handwriting}
\label{section:rw:handwriting-analysis-for-font-design}
Research into the recognition of handwritten text has led to drastic accuracy improvements~\cite{Elarian2014,Plamondon2000} and such technology can now be found in mainstream UIs (e.g., Windows, Android, iOS). However, converting digital ink into ASCII characters removes individual style.
Understanding what exactly constitutes style has been the subject of much research to inform font design~\cite{Burgert2002, Drucker1995,Noordzij05} and the related understanding of human reading has served as a source of inspiration for the modern parametric-font systems~\cite{Hussain1999,Knuth1986,Shamir1998}. Nonetheless no equivalent parametric model of handwritten style exists and analysis and description of style remains an inexact science~\cite{Noordzij05,Pulver1972}. We propose to learn a latent representation of style and to leverage it in a generative model of user specific text.

\subsection{Pen-based interaction}
Given the naturalness of the medium~\cite{Sellen2003, Cherubini2007}, pen-based interfaces have seen enduring interest in both the graphics and HCI literature~\cite{Sutherland2016}. Ever since Ivan Sutherland's Sketchpad~\cite{Sutherland1963} researchers have explored sensing and input techniques for small screens~\cite{Kienzle2013, Yoon2013}, tablets~\cite{Hinckley2014,Pfeuffer2017} and whiteboards~\cite{Mynatt1999,Perteneder2015,Xia2017} and have proposed ways of integrating paper with digital media \cite{Haller2010, Brandl2010}. Furthermore many domain specific applications have been proposed. For instance, manipulation of hand-drawn diagrams \cite{Arvo2005appearance} and geometric shapes \cite{Arvo2000fluid}, note-taking (e.g., NiCEBook~\cite{Brandl2010}), sharing of notes (e.g., NotePals~\cite{Davis1999}), browsing and annotation of multimedia content~\cite{Weibel2012}, including digital documents~\cite{Yoon2014} using a stylus. Others have explored creation, management and annotation of handwritten notes on large screen displays~\cite{Perteneder2015,Xia2017}. Typically such approaches do not convert ink into characters to preserve individual style. Our work enables new interactive possibilities by making digital ink editable and interchangeable with a character representation, allowing for advanced processing, searching and editing.

\subsection{Handwriting Beautification}
\label{section:rw:handwriting-beautification}
Zitnick~\cite{Zitnick2013} proposes a method for beautification of digital ink by exploiting the smoothing effect of geometric averaging of multiple instances of the same stroke. While generating convincing results, this method requires several samples of the same text for a single user. A supervised machine learning method to remove slope and slant from handwritten text and to normalize it's size~\cite{Espana-Boquera2011} has been proposed. Zanibbi et al \cite{Zanibbi2001aiding} introduce a tool to improve legibility of handwritten equations by applying style-preserving morphs on user-drawn symbols.
Lu et al.~\cite{Lu2012} propose to learn style features from trained artist, and to subsequently transfer the strokes of a different writer to the learnt style and therefore inherently remove the original style.
Text beautification is one of the potential applications for our work, however the proposed model only requires a single sequence as input, retains the global style of the author when beautifying and can generate novel (i.e., from ASCII characters) sequences in that style.

\subsection{Handwriting Synthesis}
\label{section:rw:handwriting-synthesis}
A large body of work is dedicated to the synthesis of handwritten text (for a comprehensive survey see ~\cite{Elarian2014}).
Attempts have been made to formalize plausible biological models of the processes underlying handwriting~\cite{Hinton2005} or by learning sequences of motion primitives~\cite{Williams2007}. Such approaches primarily validate bio-inspired models but do not produce convincing sequences. In \cite{Bhattacharya2017sigma, Plamondon2014recent} sigma-lognormal models are proposed to synthesize handwriting samples by parameterizing rapid human movements and hence reflecting writer's fine motor control capability. The model can naturally synthesize variances of a given sample, but it lacks control of the content.

Realistic handwritten characters such as Japanese Kanji or individual digits can be synthesized from learned statistical models of stroke similarity~\cite{Chang2012} or control point positions (requiring characters to be converted to splines)~\cite{Wang2002}. Follow-up work has proposed methods that connect such synthetic characters~\cite{Chen2015,Wang2005} using a ligature model. Haines et al.~\cite{Haines2016} take character-segmented images of a single author's writing and attempts to replicate the style via dynamic programming. These approaches either ignore style entirely or learn to imitate a single reference style from a large corpus of data. In contrast, our method learns first how to separate content and style and then is capable of transferring style from a single sample to arbitrary text, providing much more flexibility.

Kingma and Welling~\cite{Kingma2013} propose a variational auto-encoder architecture for manifold learning and generative modelling. The authors demonstrate synthesis of single character digits via manipulation of two, abstract latent variables but the method can not be used for conditional synthesis.
The work most closely related to ours proposes a long short-term memory recurrent (LSTM) neural network to generate complex sequences with long-range structure such as handwritten text~\cite{Graves2013}. The work demonstrates synthesis of handwritten text in specific styles limited to samples in the training set, and its model has no notion of disentangling content from style.

\section{Method}\label{sec:method}
\begin{figure}[t]
	\centering
	\includegraphics[width=0.85\columnwidth]{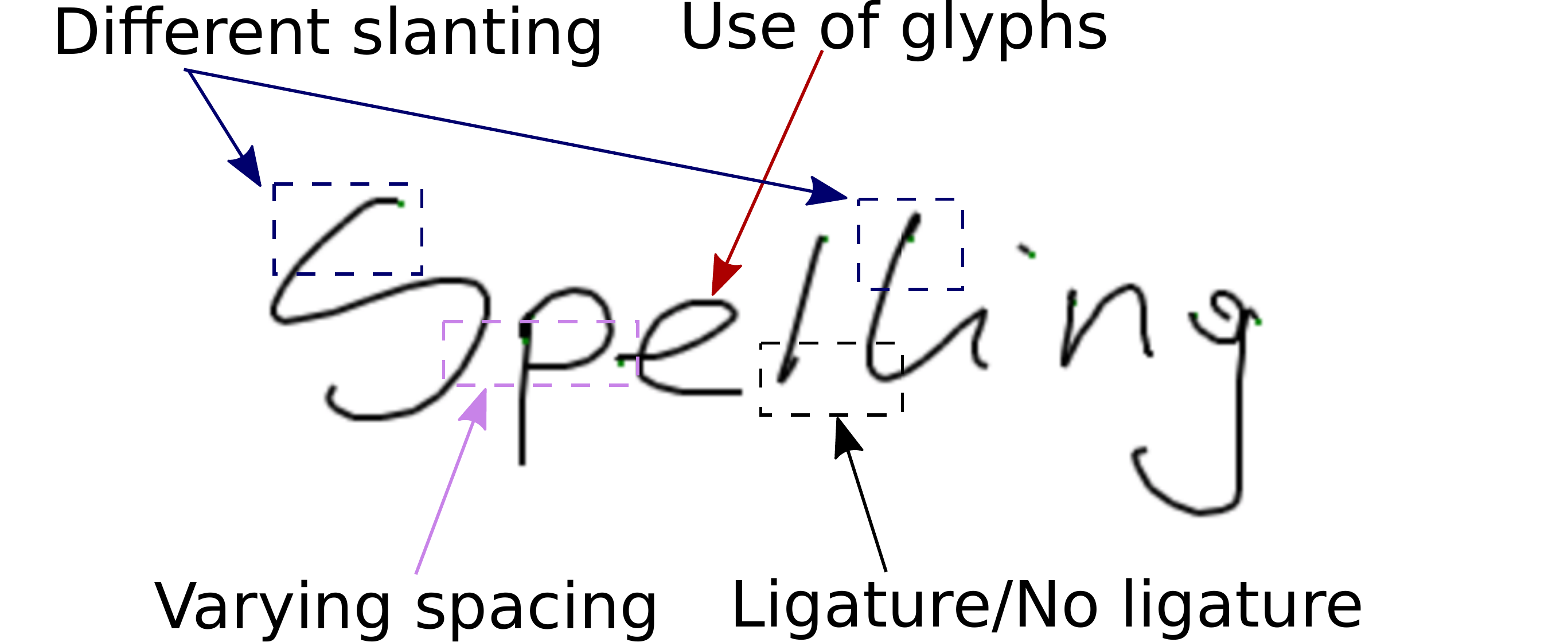}
	\caption{Example of intra-author variation that leads to entanglement of style and content, making conditional synthesis of realistic digital ink very challenging.}
	\label{fig:problem}
\end{figure}

To make digital ink fully editable, one has to overcome a number of technical problems such as character recognition and synthesis of realistic handwriting. None is more important though than the disentanglement of \textit{style} and \textit{content}. Each author has a unique style of handwriting~\cite{Srihari2002}, but at the same time, they also display a lot of intra-variability, such as mixing connected and disconnected styles, variance in usage of glyphs, character spacing and slanting (see \figref{fig:problem}). Hence, it is hard to define or predict the appearance of a character, as often its appearance is strongly influenced by its content.

\begin{figure}[t]
	\centering
	\includegraphics[width=\columnwidth, trim={20pt 230pt 360pt 20pt},clip]{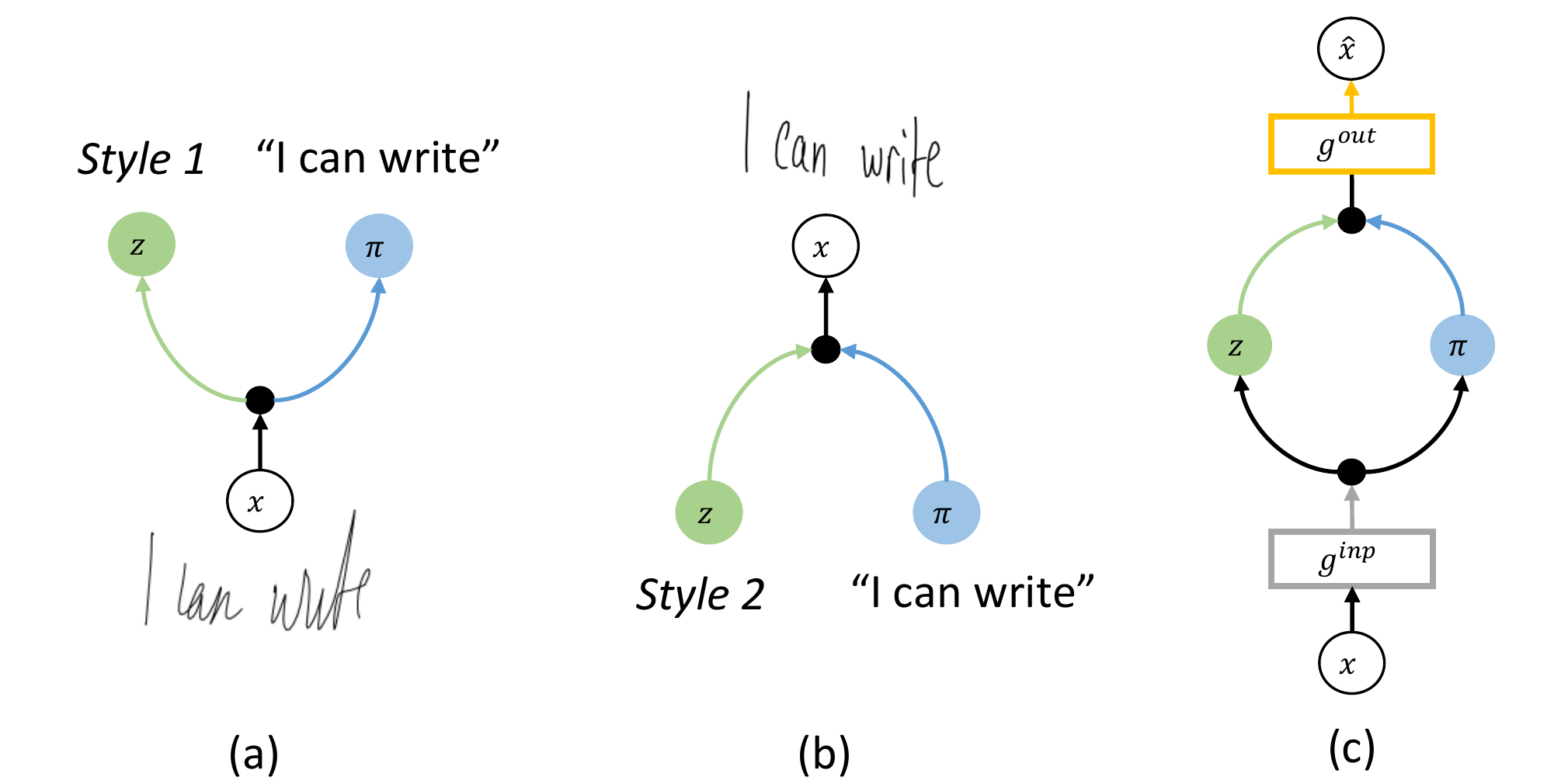}
	\caption{High-level representation of our approach. $\mathbf{x}, \mathbf{z}$ and $\mathbf{\pi}$ are random variables corresponding to handwritten text, \textit{style} and \textit{content}, respectively. (a) A given handwritten sample can be decomposed into \textit{style} and \textit{content} components. (b) Similarly, a sample can be synthesized using \textit{style} and \textit{content} components. (c) Our model learns inferring and using latent variables by reconstructing handwriting samples. $g^{inp}$ and $g^{out}$ are feed-forward networks projecting the input into an intermediate representation and predicting outputs, respectively.}
	\label{fig:factorization}
\end{figure}

In this paper, we propose a data-driven model capable of disentangling handwritten text into their content and style components, necessary to enable editing and synthesis of novel handwritten samples in a user-specified style. The key idea underlying our approach is to treat style and content as two separate latent random variables (\figref{fig:factorization}-a). While the \textit{content} component is defined as the set of alphanumeric characters and punctuation marks, the \textit{style} term is an abstraction of the factors defining appearance. It is learned by the model and projected into a continues-valued latent space. One can make use of content and style variables to \emph{edit} either style, content or both, or one can generate entirely new samples (\figref{fig:factorization}-b).

\begin{figure}[h]
	\centering
	\includegraphics[width=0.85\columnwidth, trim={25pt 400pt 680pt 20pt},clip]{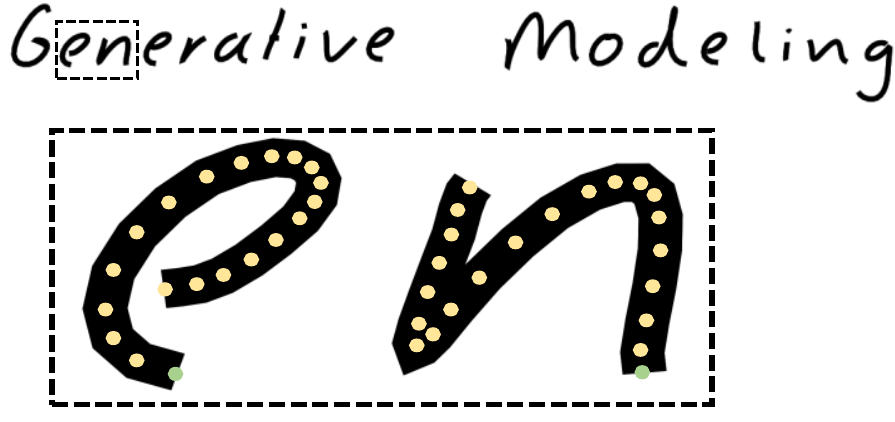}
	\caption{Discretization of digital handwriting. A handwriting sample (top) is represented by a sequence of temporarily ordered strokes. Yellow and green nodes illustrate sampled strokes. The green nodes correspond to \textit{pen-up} events.}
	\label{fig:stroke-sample}
\end{figure}

We treat digital ink as a sequence of temporarily ordered strokes where each stroke consists of $(u,v)$ pen-coordinates on the device screen and corresponding \textit{pen-up} events. While the pen-coordinates are integer values bounded by screen resolution of the device, \textit{pen-up} takes value $1$ when the pen is lifted off the screen and $0$, otherwise. A handwriting sample is formally defined as $\mathbf{x} = \{x_t\}_{t=1}^{T}$ where $x_t$ is a stroke and $T$ is total number of strokes. Moreover, we label each stroke $x_t$ with character $y_t$, end-of-character $eoc_t$ and beginning-of-word $bow_t$ labels. $y_t$ specifies which character a stroke $x_t$ belongs to. Both $eoc_t$ and $bow_t$ are binary-valued and set to $1$ if $x_t$ correspond to the last stroke of a character sequence or the first stroke of a new word, respectively (\figref{fig:stroke-sample}).

\begin{figure*}[t!]
	\centering
	\includegraphics[width=0.99\linewidth,trim={5pt 220pt 850pt 50pt},clip]{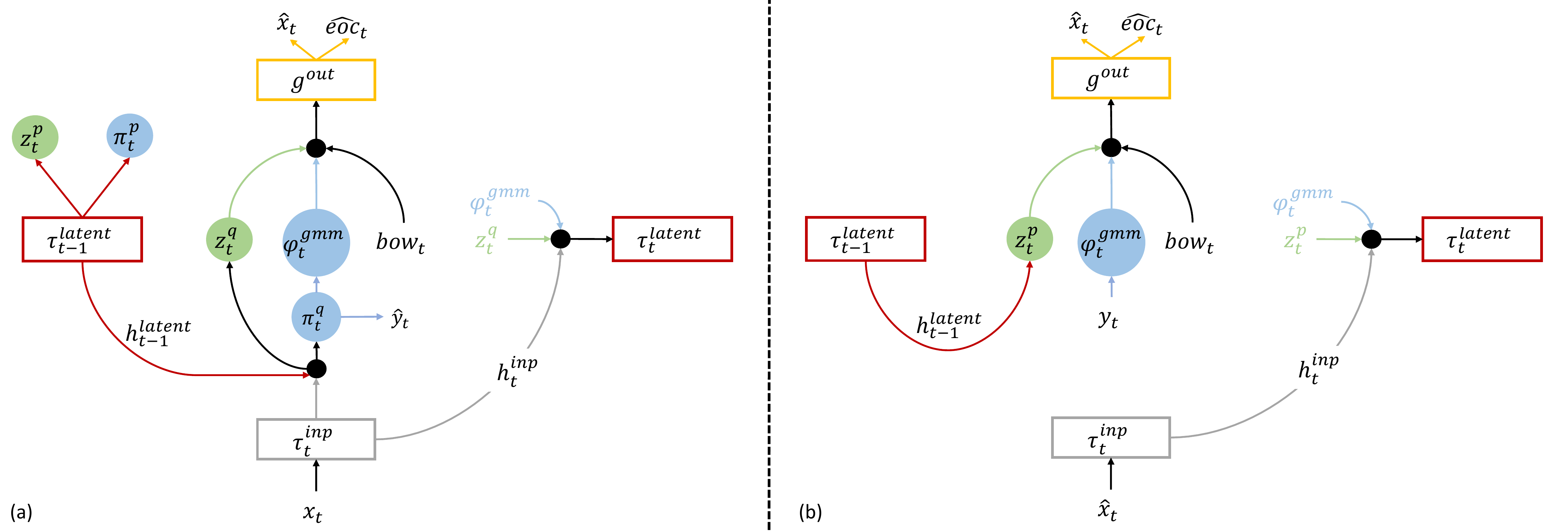}
	\caption{Schematic overview of our handwriting model in training (a) and sampling phases (b), operating at the stroke-level. Subscripts denote time-step $t$. Superscripts correspond to layer names such as \textit{input}, \textit{latent} and \textit{output} layers or the distributions of the random variables such as $z_t^q \sim q(z_t|x_t)$ and $z_t^p \sim p(z_t|x_t)$. ($\tau$ and $h$) An RNN cell and its output. ($g$) A multi-layer feed-forward neural network. (Arrows) Information flow color-coded with respect to source. (Colored circles) Latent random variables. Outgoing arrows represent a sample of the random variable. (Green branch) Gaussian latent space capturing style related information along with latent RNN ($\tau_t^{latent}$) cell at individual time-steps $t$. (Blue branch) Categorical and GMM random variables capturing content information. (Small black nodes) An auxiliary node for concatenation of incoming nodes.}
	\label{fig:model-diagram}
\end{figure*}

We propose a novel autoregressive neural network (NN) architecture that contains continuous and categorical latent random variables. Here the continuous latent variable which captures the appearance properties is modeled by an isotropic Normal distribution (\figref{fig:model-diagram} (green)). Whereas the content information is captured via a Gaussian Mixture Model (GMM), where each character in the dataset is represented by an isotropic Gaussian (shown in \figref{fig:model-diagram} (blue)). We train the model by reconstructing a given handwritten sample $\mathbf{x}$ (\figref{fig:factorization}-c). Handwriting is inherently a temporal domain and require exploiting long-range dependencies. Hence, we leverage recurrent neural network (RNN) cells and operate in the stroke level $x_t$. Moreover, we make use of and predict $y_t$, $eoc_t$, $bow_t$ in auxiliary tasks such as controlling the word spacing, character segmentation and recognition. (\figref{fig:model-diagram}).

The proposed architecture which we call conditional variational recurrent neural network (\modelname) builds on prior work on variational autoencoders (VAE) \cite{Kingma2013} and its recurrent variant, variational recurrent neural networks (VRNN) \cite{Chung2015}. While VAEs only work with non-temporal data, VRNN can reconstruct and synthesize timeseries, albeit without conditioning, providing no control over the generated content. In contrast, our model synthesizes realistic handwriting with natural variation and conditioned on a user-specified content, enabling a number of compelling applications (\figref{fig:teaser}).

\subsection{Background}
Multi-layer recurrent neural networks (RNN) \cite{Graves2013} and variational RNN (VRNN) \cite{Chung2015} are most related to our work. We briefly recap these and highlight differences. In our notation superscripts correspond to layer information such as \textit{input}, \textit{latent} or \textit{output} while subscripts denote the time-step $t$. Moreover, we drop parametrization for the sake of brevity, and therefore readers should assume that all probability distributions are modeled using neural networks.

\subsubsection{Recurrent Neural Networks}
RNNs model variable length input sequences $\mathbf{x} = (x_1, x_2, \cdots, x_T)$ by predicting the next time step $x_{t+1}$ given the current $x_t$. The probability of a sequence $\mathbf{x}$ is given by
\begin{align}
\begin{split}
\label{eq:probability_rnn}
p(\mathbf{x}) &= \prod_{t=1}^T p(x_{t+1} | x_t), \\
p(x_{t+1} | x_t) &= g^{out}(h_{t}) \\
h_t &= \tau(x_t, h_{t-1}) ,
\end{split}
\end{align}
where $\tau$ is a \emph{deterministic} transition function of an RNN cell, updating the internal cell state $h$. Note that $x_{t+1}$ implicitly depends on all inputs until step $t+1$ through the  cell state $h$.

The function $g^{out}$ maps the hidden state to a probability distribution. In vanilla RNNs (e.g., LSTM, GRU) $g^{out}$ is the only source of variability. To express the natural randomness in the data, the output function $g^{out}$ typically parametrizes a statistical distribution (e.g., Bernoulli, Normal, GMM). The output is then calculated by sampling from this distribution. Both functions $\tau$ and $g^{out}$ are approximated by optimizing neural network parameters via maximizing the log-likelihood:
\begin{equation}
\mathcal{L}_{rnn}(\mathbf{x}) = log p(\mathbf{x}) = \sum_{t=1}^T log p(x_{t+1} | x_t)
\label{eq:rnn_log_likelihood}
\end{equation}

Multi-layered LSTMs with a GMM output distribution have been used for handwriting modeling \cite{Graves2013}. While capable of conditional synthesis, they can not disentangle style from content due to the lack of latent random variables.

\subsubsection{Variational Recurrent Neural Networks} VRNNs \cite{Chung2015} modify the deterministic $\tau$ transition function by introducing a latent random variable $\mathbf{z} = (z_1, z_2, \cdots, z_T)$ increasing the expressive power of the model and to better capture variability in the data by modeling
\begin{align}
\begin{split}
\label{eq:probability_vrnn}
p(\mathbf{x}, \mathbf{z}) &= p(\mathbf{x}|\mathbf{z})p(\mathbf{z}), \\
p(\mathbf{x}, \mathbf{z}) &= \prod_{t=1}^T p(x_{t} | z_{t}) p(z_{t}), \\
p(x_{t} | z_{t}) &= g^{out}(z_t, h_{t-1}), \\
p(z_{t}) &= g^{p,z}(h_{t-1}), \\
h_t &= \tau(x_t, z_t, h_{t-1}),
\end{split}
\end{align}
where $g^{p,z}$ is a multi-layer feed forward network parameterizing the prior distribution $p(z_{t})$ and the latent variable $\mathbf{z}$ enforces the model to project the data variability on the prior distribution $p(\mathbf{z})$. Note that $x_t$ still depends on the previous steps, albeit implicitly through the internal state $h_{t-1}$.

At each time step the latent random variable $\mathbf{z}$ is modeled as isotropic Normal distribution $z_t \sim \mathcal{N}(\mu_t, \sigma_tI)$. The transition function $\tau$ takes samples $z_t$ as input, introducing a new source of variability.

Since we do not have access to the true distributions at training time, the posterior $p(\mathbf{z}|\mathbf{x})$ is intractable and hence makes the marginal likelihood, i.e., the objective, $p(\mathbf{x})$ also intractable. Instead, an approximate posterior distribution $q(\mathbf{z}|\mathbf{x})$ is employed, imitating the true posterior $p(\mathbf{z}|\mathbf{x})$ \cite{Kingma2013}, where $q(\mathbf{z}|\mathbf{x})$ is an isotropic Normal distribution and parametrized by a neural network $g^{q,z}$ as follows:
\begin{equation}
q(z_{t} | x_{t}) = g^{q,z}(x_t, h_{t-1})
\label{eq:vrnn_q}
\end{equation}
The model parameters are optimized by jointly maximizing a variational lower bound:
\begin{equation}
log p(\mathbf{x}) \geq \mathbb{E}_{q(z_t|x_t)} \sum_{t=1}^T log p(x_t | z_t) - KL(q(z_t|x_t) || p(z_t)),
\label{eq:vrnn_lower_bound}
\end{equation}

where $KL(q || p)$ is the Kullback-Leibler divergence (non-similarity) between distributions $q$ and $p$. The first term in loss (\ref{eq:vrnn_lower_bound}) ensures that the sample $x_t$ is reconstructed given the latent sample $z_t \sim q(z_{t} | x_{t})$ while the $KL$ term minimizes the discrepancy between our approximate posterior and prior distributions so that we can use the prior $z_t \sim p(z_t)$ for synthesis later. Moreover, the $q(\mathbf{z} | \mathbf{x})$ network enables inferring latent properties of a given sample, providing interesting applications. For example, a handwriting sample can be projected into the latent space $z$ and reconstructed with different slant.

A plain auto-encoder architecture learns to faithfully reconstruct input samples, the latent term $\mathbf{z}$ transforms the architecture into a fully generative model. Note that the $KL$-term never becomes $0$ due to the different amount of input information to $g^{p,z}$ and $g^{q,z}$. Hence, the $KL$-term enforces the model to capture the common information in the latent space $z$.

\subsection{Conditional Variational Recurrent Neural Network}
While multi-layer RNNs and VRNNs have appealing properties, neither is directly capable of full conditional handwriting synthesis. For example, one can synthesize a given text in a given style by using RNNs but samples will lack natural variability. Or one can generate high quality novel samples with VRNNs. However, VRNNs lack control over \emph{what} is written. Neither model have inference networks to decouple style and content, which lies at the core of our work.

We overcome this issue by introducing a new set of latent random variables, $\mathbf{z},\mathbf{\pi}$, capturing style and content of handwriting samples. More precisely our new model describes the data as being generated by two latent variables $\mathbf{z}$ and $\mathbf{\pi}$ (\figref{fig:factorization}) such that
\begin{align}
\begin{split}
\label{eq:probability_cvrnn}
p(\mathbf{x}, \mathbf{z}, \mathbf{\pi}) &= p(\mathbf{x}|\mathbf{z},\mathbf{\pi})p(\mathbf{z})p(\mathbf{\pi}), \\
p(\mathbf{x}, \mathbf{z}, \mathbf{\pi}) &= \prod_{t=1}^T p(x_{t} | z_{t}) p(z_{t}) p(\pi_{t}), \\
p(x_{t} | z_{t}, \pi_{t}) &= g^{out}(z_t, \pi_{t}), \\
p(z_{t}) &= g^{p,z}(h_{t-1}^{latent}), \\
p(\pi_{t}) &= g^{p,\pi}(h_{t-1}^{latent}), \\
h_t^{latent} &= \tau^{latent}(x_t, z_t, \pi_{t}, h_{t-1}^{latent}), \\
q(z_{t} | x_{t}) &= g^{q,z}(x_t, h_{t-1}^{latent}),
\end{split}
\end{align}

where $p(\pi_t)$ is a $K$-dimensional multinomial distribution specifying the characters that are synthesized.

Similar to VRNNs, we introduce an approximate inference distribution $q(\mathbf{\pi} | \mathbf{x})$ for the categorical latent variable:
\begin{equation}
\begin{split}
q(\pi_t | x_{t}) &= g^{q,\pi}(x_t, h_{t-1}^{latent})
\label{eq:cvrnn_pi_q}
\end{split}
\end{equation}
Since we aim to decouple style and content in handwriting, we assume that the approximate distribution has a factorized form $q(z_t, \pi_t|x_t) = q(z_t|x_t)q(\pi_t|x_t)$. Both $q(\mathbf{\pi} | \mathbf{x})$ and $q(\mathbf{z} | \mathbf{x})$ are used to infer content and style components of a given sample $\mathbf{x}$ as described earlier.

We optimize the following variational lower bound:
\begin{align}
\begin{split}
log p(\mathbf{x}) \geq \mathcal{L}_{lb}(\cdot) = \mathbb{E}_{q(z_t, \pi_t|x_t)} \sum_{t=1}^T log p(x_t | z_t, \pi_t) \\ - KL(q(z_t|x_t) || p(z_t)) - KL(q(\pi_t|x_t) || p(\pi_t)),
\label{eq:cvrnn_lower_bound}
\end{split}
\end{align}
where the first term ensures that the input stroke is reconstructed by using its latent samples. We model the output by using bivariate Gaussian and Bernoulli distributions for $2D$-pixel coordinates and binary \textit{pen-up} events, respectively.

Note that our output function $g^{out}$ does not employ the internal cell state $h$. By using only the latent variables $z$ and $\pi$ for synthesis, we aim to enforce the model to capture the patterns only in the latent variables $z$ and $\pi$.

\subsection{High Quality Digital Ink Synthesis}
The \modelname architecture as discussed so far enables the crucial component of separating continuous components from categorical aspects (i.e., characters) which potentially would be sufficient to conditionally synthesize individual characters. However, to fully address the entire handwriting task several extension to control important aspects such as word-spacing and to improve quality of the predictions are necessary.

\subsubsection{Character classification loss} Although we assume that the latent random variables $\mathbf{z}$ and $\mathbf{\pi}$ capture style and content information, respectively, and make a conditional independence assumption, in practice full disentanglement is an ambiguous task. Since we essentially ask the model to learn by itself what style and what content are we found  further guidance at training time to be necessary. 

To prevent divergence during training we make use of character labels at training time and add an additional cross-entropy classification loss $\mathcal{L}_{classification}$ on the content component $q(\pi_t|x_t)$.

\subsubsection{GMM latent space} 
\begin{figure}[t]
	\centering
	\includegraphics[width=1\columnwidth]{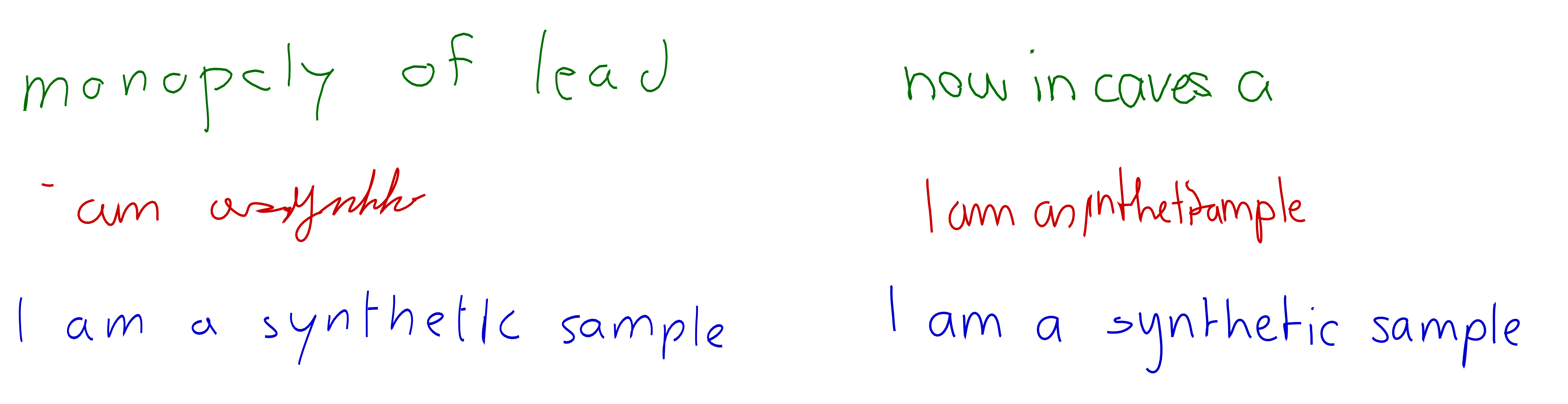}
	\caption{(top, green) Input samples used to infer style. (middle, red) Synthetic samples of a model with $\mathbf{\pi}$ only. They are generated using one-hot-encoded character labels, causing problems with \textit{pen-up} events and with character placement. (bottom, blue) Synthetic samples of our model \emph{with} GMM latent space.}
	\label{fig:gmm-vs-one-hot}
\end{figure}

Conditioning generative models is typically done via one-hot encoded labels. While we could directly use samples from $q(\pi_t|x_t)$, we prefer using a continuous representation. We hypothesize and experimentally validate (see \figref{fig:gmm-vs-one-hot}) that the synthesis model can shape the latent space with respect to the loss caused by the content aspect. 

\begin{figure}[t]
	\centering
	\includegraphics[width=0.85\columnwidth]{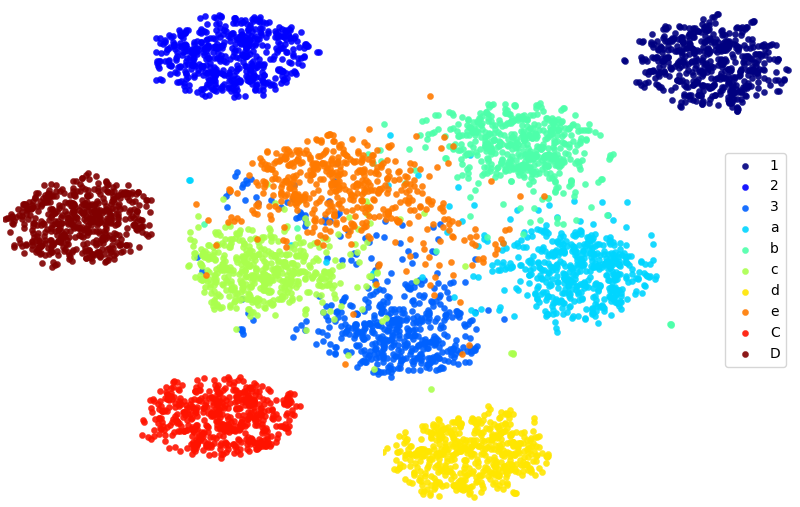}
	\caption{Illustration of our GMM latent space $\varphi^{gmm}$. We select a small subset of our alphabet and draw $500$ samples from corresponding GMM components. We use the t-sne algorithm \protect\cite{maaten2008visualizing} to visualize $32$-dimensional samples in $2$D space. Note that the t-sne algorithm finds an arbitrary placement and hence the positioning does not reflect the true latent space. Nevertheless, letters form separate clusters.}
	\label{fig:gmm_latent_space}
\end{figure}

For this purpose we use a Gaussian mixture model where each character in $K$ is represented by an isotropic Gaussian
\begin{align}
\begin{split}
p(\varphi_t) = \sum_{k=1}^K \pi_{t,k} \mathcal{N}(\varphi_t | \mu_k, \sigma_k)   ,
\label{eq:gmm_latent}
\end{split}
\end{align}
where $\mathcal{N}(\varphi_t | \mu_k, \sigma_k)$ is the probability of sampling from the corresponding mixture component $k$. $\pi$ corresponds to the content variable in \refequ{eq:cvrnn_pi_q} which is here interpreted as weight of the mixture components. This means that we use $q(\pi_t|x_t)$ to select a particular Gaussian component  for a given stroke sample $x_t$. We then sample $\varphi_t$ from the $k$-th Gaussian component and apply the ``re-parametrization trick'' \cite{Kingma2013,Gurumurthy2017} so that the gradients can flow through the random variables, enabling the learning of GMM parameters via standard backpropagation.
\begin{align}
\begin{split}
\varphi_t = \mu_k + \sigma_k\epsilon,
\label{eq:gmm_reparametrization_trick}
\end{split}
\end{align}
where $\epsilon \sim \mathcal{N}(0,1)$. Our continuous content representation results in similar letters being located closer in the latent space while dissimilar letters or infrequent symbols being pushed away. This effect is visualized in \figref{fig:gmm_latent_space}.

Importantly the GMM parameters are sampled from a time-invariant distribution. That  is they remain the same for all data samples and across time steps of a given input $\mathbf{x}$, whereas $z_t$ is dynamic and employs new parameters per time step. For each Gaussian component in $\varphi$, we initialize $\mu_k$, $1 \leq k \leq K$, randomly by using a uniform distribution $\mathcal{U}(-1,1)$ and $\sigma_k$ with a fixed value of $1$. The GMM components are trained alongside the other network parameters. 

In order to increase model convergence speed and to improve results, we use ground truth character labels during training. More precisely, the GMM components are selected by using the real labels $y$ instead of predictions of the inference network $q(\pi_t|x_t)$. Instead $q(\pi_t|x_t)$ is trained only by using the classification loss $\mathcal{L}_{classification}$ and not affected by the gradients of GMM with respect to $\pi_t$.

\subsubsection{Word spacing and character limits}
At sampling time the model needs to automatically infer word spacing and which character to synthesize (these are a priori unknown). In order to control when to leave a space between words or when to start synthesizing the next character, we introduce two additional signals during training, namely \textit{eoc} and \textit{bow} signaling the \emph{end} of a character and \emph{beginning} of a word respectively. These labels are attained from ground truth character level segmentations (see Dataset section).

The \textit{bow} signal is fed as input to the output function $g^{out}$ and the output distribution of our handwriting synthesis takes the following form:
\begin{align}
\begin{split}
p(x_{t} | z_{t}, \pi_{t}) &= g^{out}(z_t, \varphi_t, bow_t),
\label{eq:z_sow_input}
\end{split}
\end{align}
forcing the model to learn when to leave empty space at training and sampling time.

The \textit{eoc} signal, on the other hand, is provided to the model at training so that it can predict when to stop synthesizing a given character. It is included in the loss function in the form of Bernoulli log-likelihood $\mathcal{L}_{eoc}$. Along with the reconstruction of the input stroke $x_t$, the $eoc_t$ label is predicted.

\begin{algorithm}[t]
	\caption{Training \\ The flow of information through the model at training time, from inputs to outputs. The model is presented in \figref{fig:model-diagram}-b color coded components in comments.} \label{alg:training}
	\begin{algorithmic}[1]
		\Require{}
		\Statex Strokes $\mathbf{x} = \{x_t\}_{t=1}^{T}$
		\Statex Labels ($\mathbf{y, eoc, bow}) = \{(y_t, eoc_t, bow_t)\}_{t=1}^{T}$
		\Statex $h_0^{inp} = h_0^{latent} = \mathbf{0}$
		\Ensure{}
		\Statex Reconstructed strokes $\mathbf{\hat{x}}$, predicted $\mathbf{\hat{y}}$ and $\mathbf{\hat{eoc}}$ labels.
		\State $t \gets 1$
		\While{$t= \le T$}
		\State \parbox[t]{\dimexpr\linewidth-\algorithmicindent}{Update $h_t^{inp} \gets \tau^{inp}(x_t, h_{t-1}^{inp})$ \Comment{grey} \strut}
		\State \parbox[t]{\dimexpr\linewidth-\algorithmicindent}{Sample $\pi_t^q \sim q(\pi_t | x_{t})$ using \refequ{eq:pi_t_q} (equivalent to the character prediction $\hat{y}_t$), \Comment{blue} \strut}
		\State \parbox[t]{\dimexpr\linewidth-\algorithmicindent}{Select the corresponding GMM component and draw a content sample $\varphi_t$ using \refequ{eq:gmm_reparametrization_trick}, \Comment{blue} \strut}
		\State \parbox[t]{\dimexpr\linewidth-\algorithmicindent}{Estimate parameters of the isotropic Gaussian $q(z_{t} | x_{t})$ using \refequ{eq:z_t_q} and sample $z_t^q \sim q(z_{t} | x_{t})$, \Comment{green} \strut}
		\State \parbox[t]{\dimexpr\linewidth-\algorithmicindent}{Using $\varphi_t$, $z_t^q$ and $bow_t$, reconstruct the stroke $\hat{x_t}$ and predict $\widehat{eoc_t}$, \Comment{yellow}\strut}
		\State \parbox[t]{\dimexpr\linewidth-\algorithmicindent}{Estimate isotropic Gaussian and Multinomial distribution parameters of prior latent variables $z_t^p$ and $\pi_t^p$ by using \refequs{eq:z_t_p}{eq:pi_t_p}, respectively, \strut}
		\State \parbox[t]{\dimexpr\linewidth-\algorithmicindent}{Update $h_t^{latent} \gets \tau^{latent}(h_t^{inp}, z_t^q, \varphi_t, h_{t-1}^{latent})$, \strut}
		\State $t \gets t + 1$
		\EndWhile
		\State \parbox[t]{\dimexpr\linewidth-\algorithmicindent}{Evaluate \refequ{eq:total_loss} and update model parameters. \strut}
	\end{algorithmic}
\end{algorithm}
\vspace{-0em}
\subsubsection{Input RNN cell}
Our model consists of two LSTM cells in the latent and at the input layers. Note that the latent cell is originally contributing via the transition function $\tau^{latent}$ in \refequ{eq:probability_cvrnn}. Using an additional cell at the input layer increases model capacity (similar to multi-layered RNNs) and adds a new transition function $\tau^{inp}$. Thus the synthesis model can capture and modulate temporal patterns at the input levels. Intuitively this is motivated by the strong temporal consistency in handwriting where the previous letter influences the appearance of the current (cf. \figref{fig:problem}).

We now use a temporal representation $h_t^{inp}$ of the input strokes $x_t$. With the cumulative modifications, our \modelname architecture becomes
\begin{align}
\label{eq:x_generation}
p(x_{t} | z_{t}, \pi_{t}) &= g^{out}(z_t, \varphi_t, bow_t), \\
\label{eq:z_t_p}
z_t^p \sim p(z_{t}) &= g^{p,z}(h_{t-1}^{latent}), \\
\label{eq:pi_t_p}
\pi_{t}^p \sim p(\pi_{t}) &= g^{p,\pi}(h_{t-1}^{latent}), \\
\label{eq:h_t_inp}
h_t^{inp} &= \tau^{inp}(x_t, h_{t-1}^{inp}), \\
\label{eq:z_t_q}
z_t^q \sim q(z_{t} | x_{t}) &= g^{q,z}(h_t^{inp}, h_{t-1}^{latent}), \\
\label{eq:pi_t_q}
\pi_t^q \sim q(\pi_t | x_{t}) &= g^{q,\pi}(h_t^{inp}, h_{t-1}^{latent}), \\
\label{eq:h_t_latent}
h_t^{latent} &= \tau^{latent}(h_t^{inp}, z_t, \varphi_t, h_{t-1}^{latent}).
\end{align}

Finally we train our handwriting model by using algorithm (\ref{alg:training}) and optimizing the following loss:
\begin{align}
\begin{split}
\mathcal{L}(\cdot) = \mathcal{L}_{lb} + \mathcal{L}_{classification} + \mathcal{L}_{eoc}.
\label{eq:total_loss}
\end{split}
\end{align}

In our style transfer applications, by following the steps $1-11$ of algorithm (\ref{alg:training}), we first feed the model with a reference sample and get the internal state of the latent LSTM cell $h^{latent}$ carrying style information. The sampling algorithm (\ref{alg:sampling}) to generate new samples is then initialized with this $h^{latent}$.

We implement our model in Tensorflow \cite{Abadi2016tensorflow}. The model and training details are provided in the Appendix. Code and dataset for learning and reproducing our results can be found at \url{https://ait.ethz.ch/projects/2018/deepwriting/}.

\begin{algorithm}[t]
	\caption{Sampling \\ The procedure to \textit{synthesize} novel handwriting sequences by conditioning on content and style. The model is presented in \figref{fig:model-diagram}-c and color coded components in comments.} \label{alg:sampling}
	\begin{algorithmic}[1]
		\Require{}
		\Statex Sequence of characters $\mathbf{y} = \{y_n\}_{n=1}^{N}$ to be synthesized.
		\Statex $h_0^{latent}$ is zero-initialized or inferred from another sample.
		\Statex Probability threshold $\epsilon$ for switching to the next character.
		\Ensure{}
		\Statex Generated strokes $\mathbf{\hat{x}}$ of the given text $\mathbf{y}$.
		\State $t \gets 1$, $n \gets 1$.
		\While{$n \le N$}
		\State \parbox[t]{\dimexpr\linewidth-\algorithmicindent}{Estimate parameters of style prior $p(z_t)$ using \refequ{eq:z_t_p} and sample $z_t^p$. \Comment{green} \strut}
		\State \parbox[t]{\dimexpr\linewidth-\algorithmicindent}{$\pi_t^p \gets y_n$, select the corresponding GMM component and draw a content sample. $\varphi_t$ using \refequ{eq:gmm_reparametrization_trick}, \Comment{blue} \strut}
		\State \parbox[t]{\dimexpr\linewidth-\algorithmicindent}{$bow_t \gets 1$ if it is the first stroke of a new word, $0$, otherwise. \strut}
		\State \parbox[t]{\dimexpr\linewidth-\algorithmicindent}{Using $\varphi_t$, $z_t^p$ and $bow_t$, synthesize a new stroke $\hat{x_t}$ and predict $\widehat{eoc_t}$, \Comment{yellow}\strut}
		\State \parbox[t]{\dimexpr\linewidth-\algorithmicindent}{$n \gets n + 1$ if $\widehat{eoc}_t > \epsilon$, \Comment{next character} \strut}
		\State $t \gets t + 1$
		\EndWhile
	\end{algorithmic}
\end{algorithm}
\vspace{-0em}
\subsection{Character recognition}
Disentanglement of content requires character recognition of handwritten input. Our model uses the latent variable $\mathbf{\pi}$ to infer character labels which is fed by the input LSTM cell. In our experiments we observe that bidirectional recurrent neural networks (BiRNN) \cite{schuster1997bidirectional} perform significantly better than standard LSTM models in content recognition task ($60\%$ against $96\%$ validation accuracy). BiRNNs have access to future steps by processing the input sequence from both directions. However, they also require that the entire input sequence must be available at once. This inherent constraint makes it difficult to fully integrate them into our training and sampling architecture where the input data is available one step at a time.

Due to their desirable performance we train a separate BiRNN model to classify input samples. At sampling time we use its predictions to guide disentanglement. Note that technically $q(\mathbf{\pi} | \mathbf{x})$ also infers character labels but the BiRNN's accuracy improves the quality of synthesis. We leave full integration of high accuracy character recognition for future work.

\section{Application Scenarios}\label{sec:apps}
By disentangling content from style, our approach makes digital ink truly editable. This allows the generation of novel writing in user-defined styles and, similarly to typed text, of seamless editing of handwritten text. Further, it enables a wide range of exciting application scenarios, of which we discuss proof-of-concept implementations.

\subsection{Conditional Handwriting Generation}

\begin{figure}[h!]
\centering
\includegraphics[width=\columnwidth]{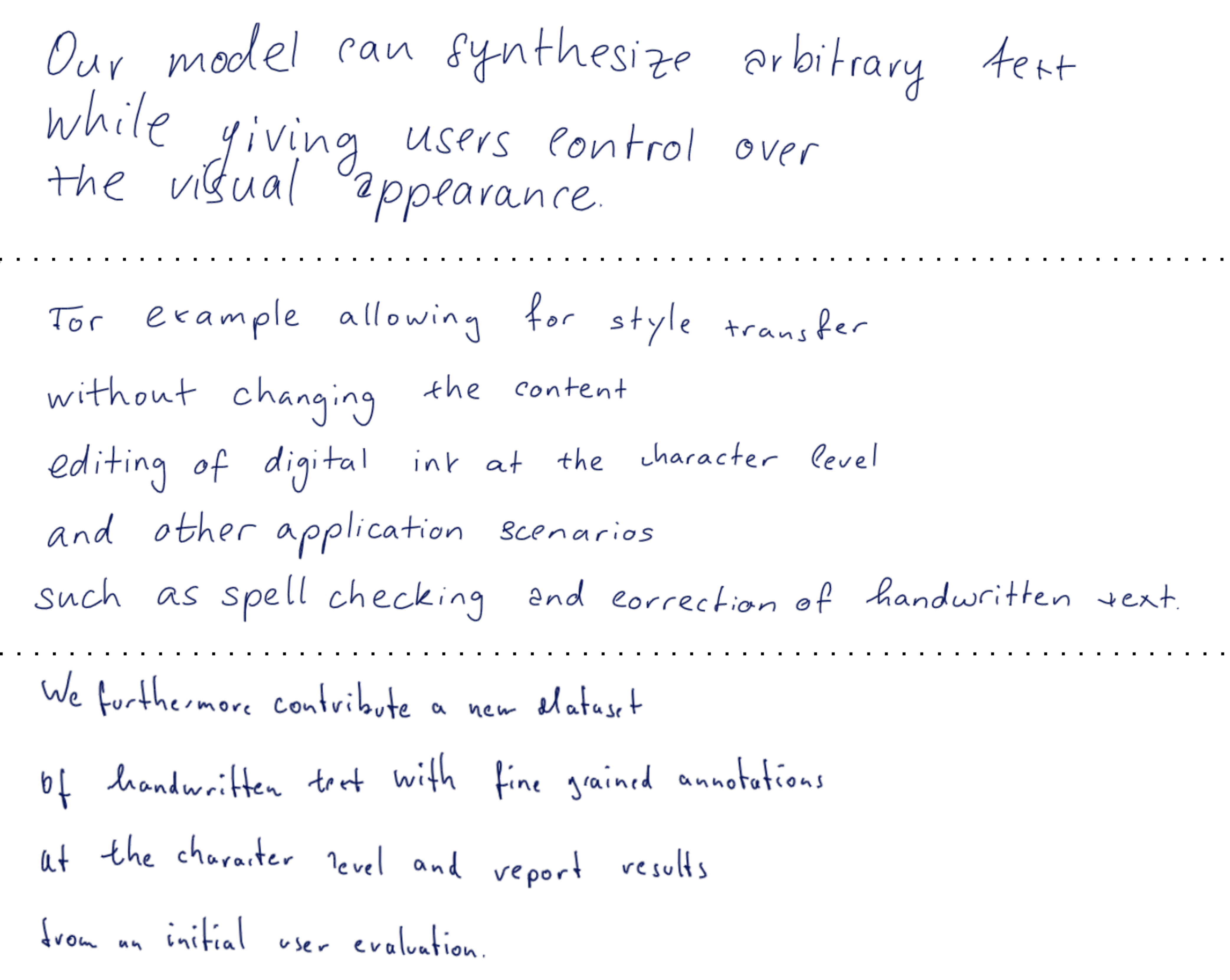}
\caption{Handwritten text synthesized from the paper abstract. Each sentence is ``written'' in the style of a different author. For full abstract, see Appendix.}
\label{fig:interface-synthesis}
\end{figure}

To illustrate the capability to synthesize \emph{novel} text in a user-specific style we have implemented an interface that allows user to type text, browse a database of handwritten samples from different authors and to generate novel handwriting. The novel sequence takes the content from the typed text and matches the style to a single input sample (cf. video). This could be directly embedded in existing note-taking applications to generate personalized handwritten notes from typed text or email clients could turn typed text into handwritten, personalized letters. For demonstration we have synthesized extracts of this paper's abstract in three different styles (see~\figref{fig:interface-synthesis}).

\subsection{Content Preserving Style Transfer}

\begin{figure}[th]
\centering
\includegraphics[width=0.8\columnwidth,trim={0pt 950pt 0pt 0pt},clip]{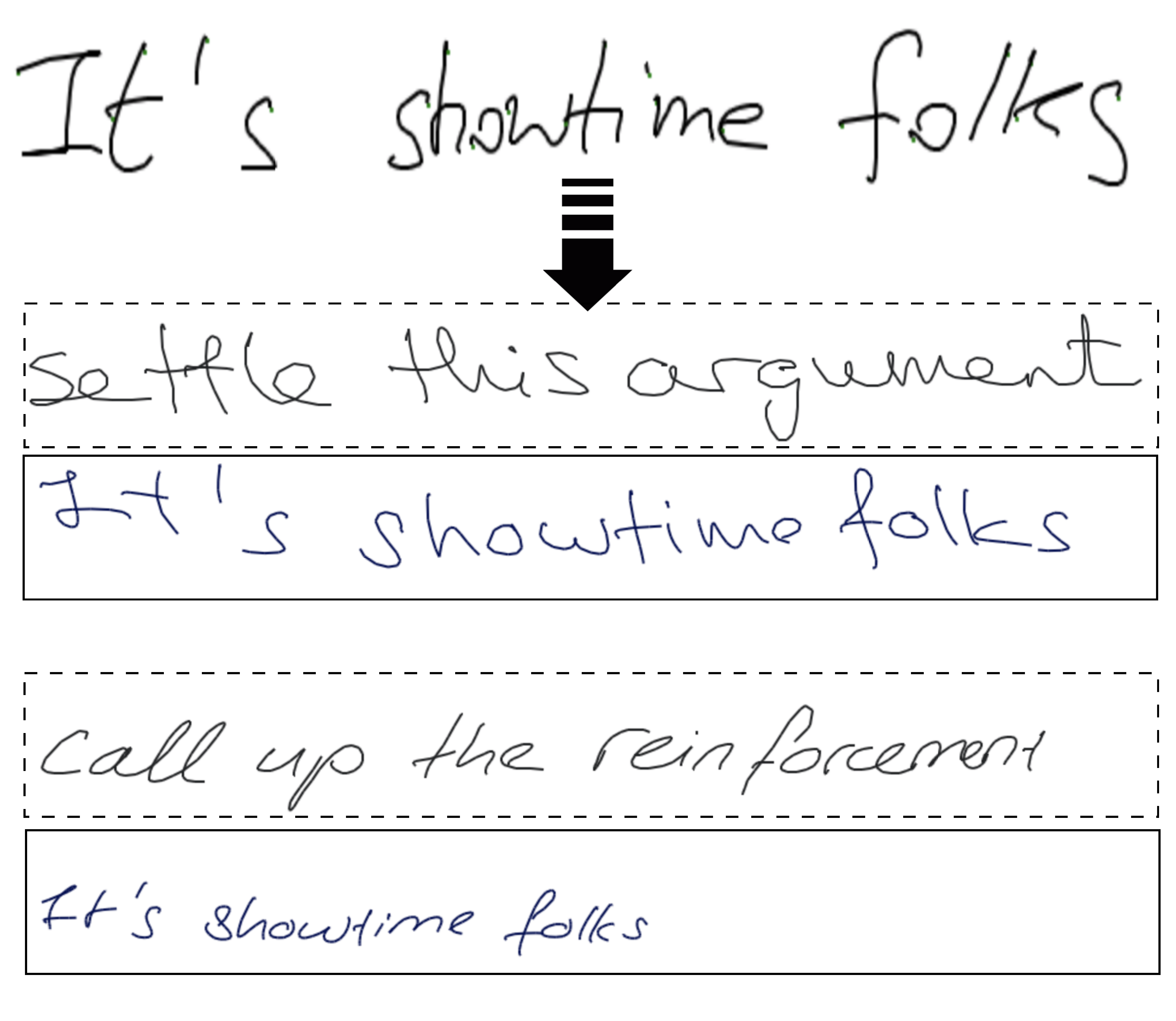}
\caption{Style transfer. The input sequence (top) is transferred to a selected reference style (black ink, dotted outlines). The results (blue ink, solid outline) preserve the input content, and its appearance matches the reference style.}
\label{fig:interface-transfer}
\vspace{-7pt}
\end{figure}

Our model can furthermore transfer \emph{existing} handwritten samples to \emph{novel} styles, thus preserving their content while changing their appearance. We implemented an interactive tablet application that allows users to recast their own handwriting into a selected style (see \figref{fig:interface-transfer} for results and video figure for interface walk through). After scribbling on the canvas and selecting an author's handwriting sample, users see their strokes morphed to that style in real-time. Such solution could be beneficial for a variety of domains. For example, artist and comic authors could include specific handwritten lettering in their work, or preserving style during localization to a foreign language.

\paragraph{Beautification}
When using the users own input style as target style, our model re-generates smoother versions of the original strokes, while maintaining natural variability and diversity. Thus obtaining an averaging effect that suppresses local noise and preserves global style features. The resulting strokes are then beautified (see \figref{fig:exp_result} and video), in line with previous work that solely relied on token averaging for beautification (e.g., \cite{Zitnick2013}) or denoising (e.g., \cite{Buades2005}).

\subsection{Word-level Editing}

\begin{figure}[h]
	\centering
	\includegraphics[width=0.8\columnwidth]{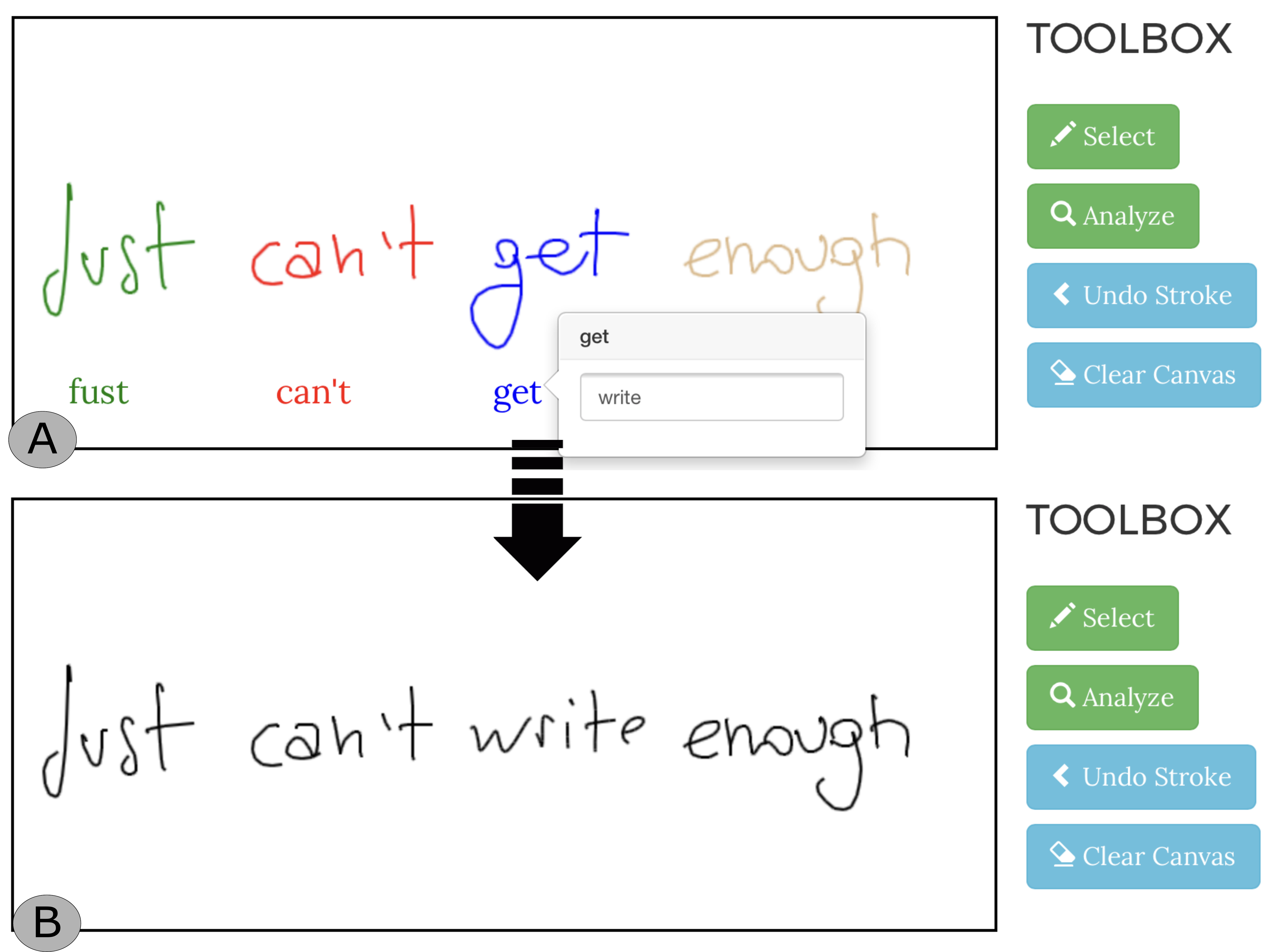}
	\caption{Our model allows editing of handwritten text at the word level. A) Handwriting is recognized, with each individual word fully editable. B) Edited words are synthesized and embedded in the original text, preserving the style.}
	\label{fig:edit}
\end{figure}

\begin{figure}[h]
	\centering
	\includegraphics[width=0.9\columnwidth]{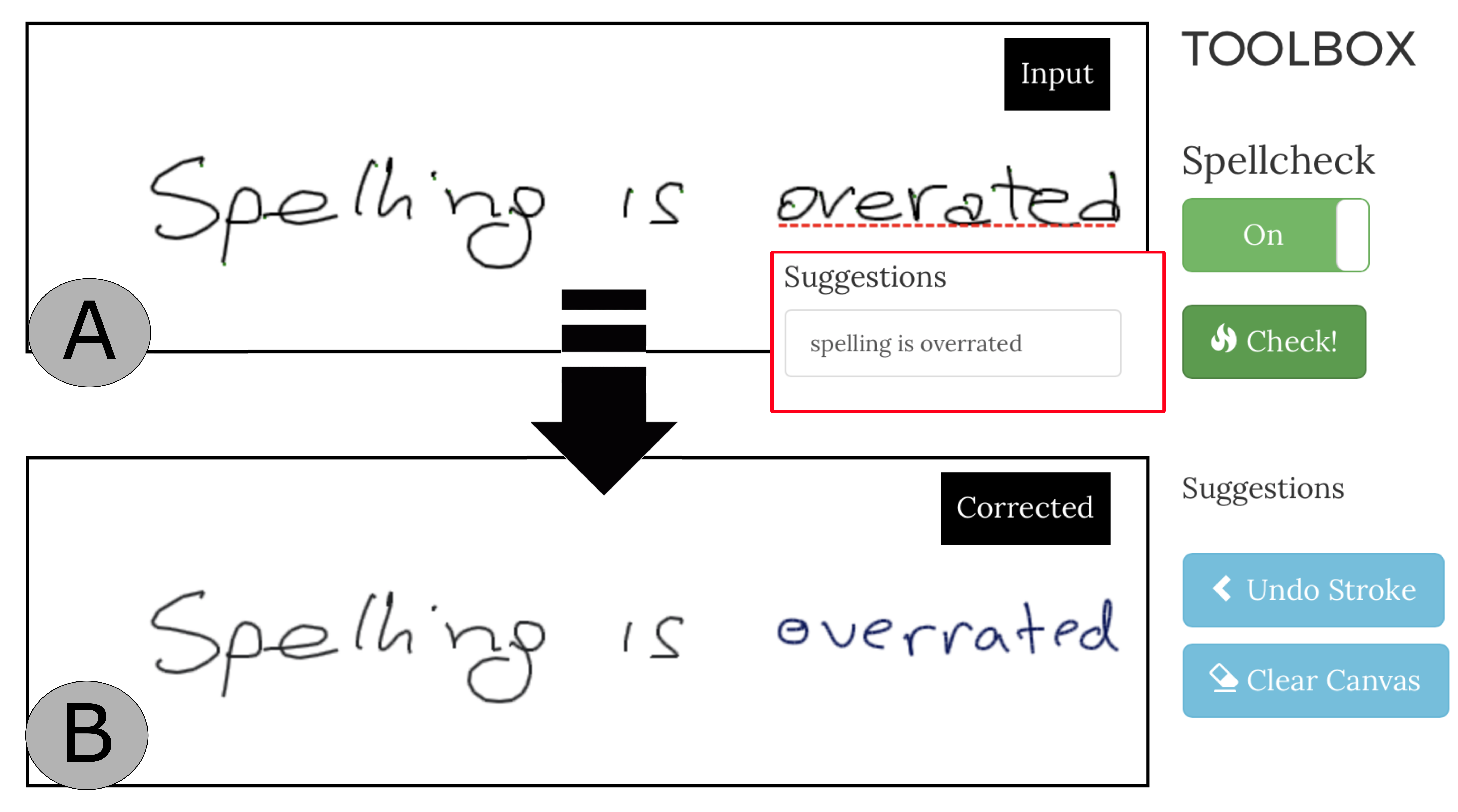}
	\caption{Our spell-checking interface. A) Spelling and grammar mistakes are detected and highlighted directly on the handwriting. Alternative spelling is offered (red box). B) Corrected words are synthesized and embedded in the original text (blue ink), preserving the writer style. }
	\label{fig:interface-spellcheck}
\end{figure}

At the core of our technique lies the ability to edit digital ink at the same level of fidelity as typed text, allowing users to change, delete or replace individual words. \figref{fig:edit} illustrates a simple prototype allowing users to edit handwritten content, while preserving the original style when re-synthesizing it.
Our model recognizes individual words and characters and renders them as (editable) overlays. The user may select individual words, change the content, and regenerate the digital ink reflecting the edits while maintaining a coherent visual appearance. We see many applications, for example note taking apps, which require frequent edits but currently do not allow for this without loosing visual appearance. 

\paragraph{Handwriting Spell-checking and Correction}

A further application of the ability to edit digital ink at the word level is the possibility to spell-check and correct handwritten text. As a proof of concept, we implemented a functional handwriting spell-checker that can analyze digital ink, detect spelling mistakes, and correct the written samples by synthesizing the corrected sentence in the original style (see \figref{fig:interface-spellcheck} and video figure). For the implementation we rely on existing spell-checking APIs, feeding recognized characters into it and re-rendering the retrieved corrections.


\section{Preliminary User Evaluation}\label{sec:evaluation}

So far we have introduced our neural network architecture and have evaluated its capability to synthesize digital ink. We now shift our focus on initially evaluating users' perception and the usability of our method. To this end, we conducted a preliminary user study gathering quantitative and qualitative data on two separate tasks. Throughout the experiment, 10 subjects ($M=27.9$; $SD=3.34$; 3 female) from our institution evaluated our model using an iPad Pro and Apple Pencil.

\begin{figure}[t]
	\centering
	\includegraphics[width=\columnwidth]{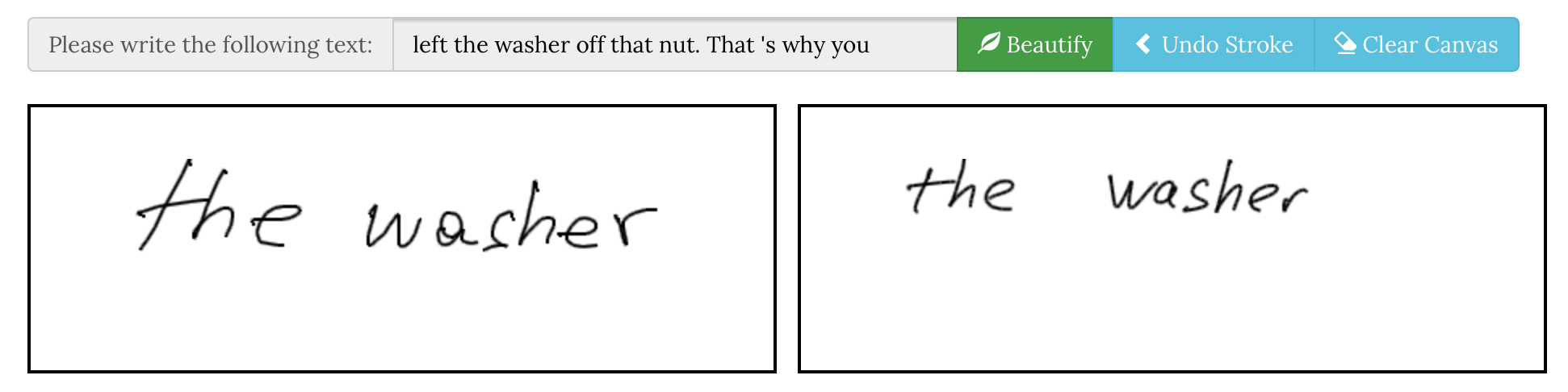} \\
	\includegraphics[width=\columnwidth]{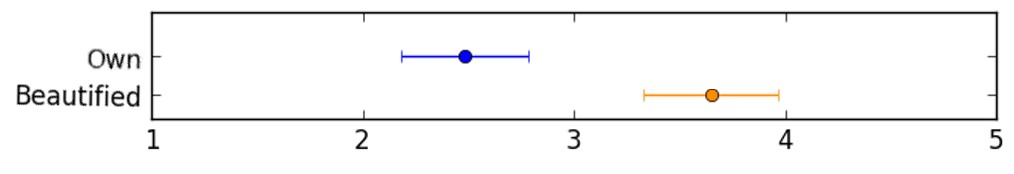}
	\caption{Task 1. Top: Experimental Interface. Participants input on the left; beautified version shown on the right. Bottom: Confidence interval plot on a 5-point Likert scale.}
	\label{fig:exp_result}
\end{figure}

\paragraph{Handwriting Beautification}
The first part of our experiment evaluates text beautification. Users were asked to compare their original handwriting with its beautified counterpart. Specifically, we asked our subjects to repeatedly write extracts from the LOB corpus \cite{Stig1986}, for a total of 12 trials each. In each trial the participant copied down the sample and we beautified the strokes with the results being shown side-by-side (see \figref{fig:exp_result}, top). Users were then asked to rate the aesthetics of their own script (Q: \textit{I find my own handwriting aesthetically pleasing}) and the beautified version (Q: \textit{I find the beautified handwriting aesthetically pleasing}), using a 5-point Likert scale. Importantly these were treated as independent questions (i.e., users were allowed to like both).

\paragraph{Handwriting Spell-Checking}
In the second task we evaluate the spell-checking utility (see \figref{fig:interface-spellcheck}). We randomly sampled from the LOB corpus and perturbed individual words such that they contained spelling mistakes. Participants then used our tool to correct the written text (while maintaining it's style), and subsequently were asked to fill in a standard System Usability Scale (SUS) questionnaire and take part in an exit interview.

\paragraph{Results}

Our results, summarized in~\figref{fig:exp_result} (bottom), indicate that users' reception of our technique is overall positive. The beautified strokes were on average rated higher ($M=3.65$, 95\% CI [3.33-3.97]) with non overlapping confidence intervals. The SUS results further supports this trend, with our system scoring positively ($SUS=85$). Following the analysis technique suggested in~\cite{Lewis2009}, our system can be classified as Rank A, indicating that users are likely to recommend it to others.

The above results are also echoed by participants' comments during the exit interviews (e.g., \textit{I have never seen anything like this}, and \textit{Finally others can read my notes.}). Furthermore, some suggested additional applications that would naturally fit our model capabilities (e.g., \textit{This would be very useful to correct bad or illegible handwriting}, \textit{I can see this used a lot in education, especially when teaching how to write to kids} and \textit{This would be perfect for note taking, as one could go back in their notes and remove mistakes, abbreviations and so on}). Interestingly, the ability to preserve style while editing content were mentioned frequently as the most valued feature of our approach (e.g., \textit{Having a spell-checker for my own handwriting feels like writing personalized text messages!}).


\section{The Handwriting Dataset}\label{sec:dataset}
Machine learning models such as ours rely on the availability of annotated training data. Prior work mostly relied on the IAM On-Line Handwriting Database (IAM-OnDB) \cite{Liwicki2005}. However, this was captured with a low-resolution and low-accuracy digital whiteboard and only contains annotations at the sequence level. To disentangle content and style, more fine-grained annotations are needed. Hence, we contribute a novel dataset of handwritten text with character level annotations. 

\begin{figure}[t]
\centering
\includegraphics[width=.8\columnwidth]{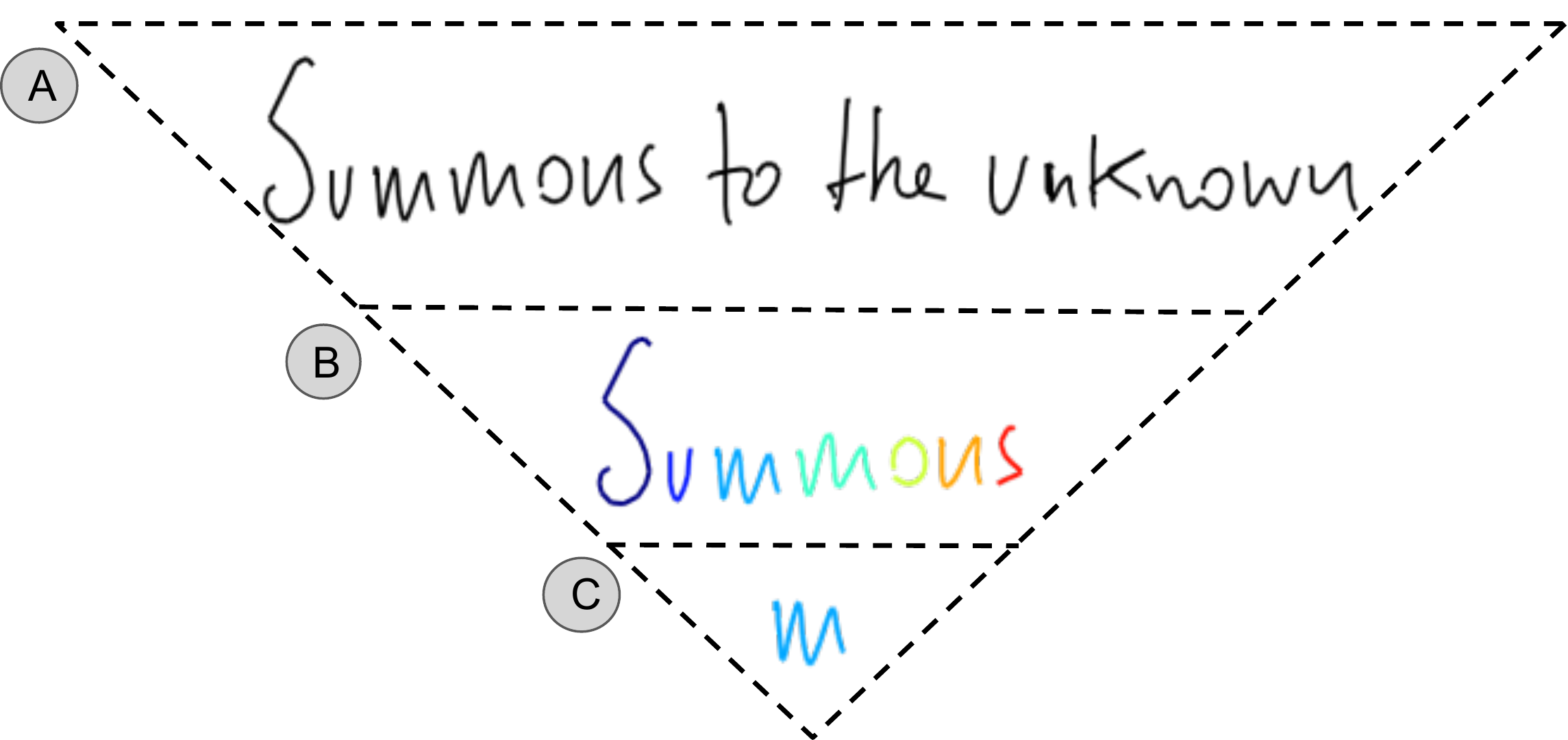}\\ \hspace{.5cm}
\includegraphics[width=.8\columnwidth]{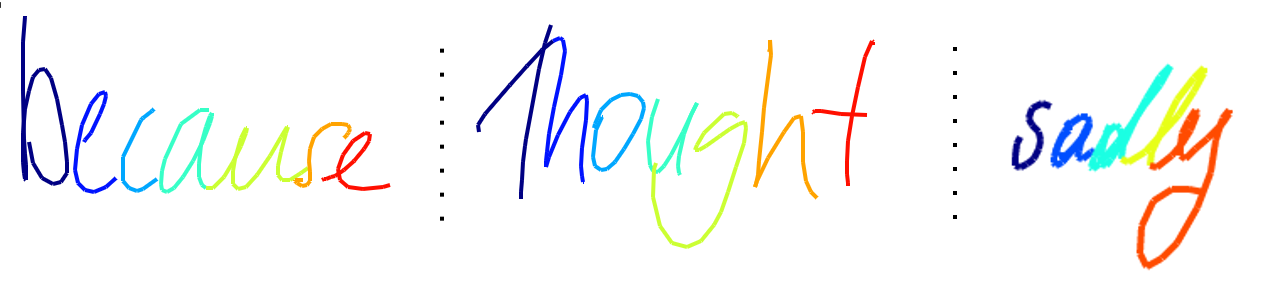}
\caption{ (Top) Our dataset offers different level of annotation, including sentence-wide annotation (A), as well as fine-grained segmentation at word (B) and character (C) level.
(Bottom) Samples from our dataset, with colour-coded character segmentation. Different styles are available, including challenging styles to segment (e.g., joined-up cursive, right).}
\label{fig:dataset}
\end{figure}

The proposed dataset accumulates the IAM-OnDB dataset with newly collected samples. At time of writing, the unified dataset contains data from 294 unique authors, for a total of 85181 word instances (writer median 292 words) and 406956 handwritten characters (writer median 1349 characters), with further statistics reported in Table~\ref{table:dataset_stats}. The data is stored using the JSON data-interchange format, which makes it easy to distribute and use in a variety of programming environments.

\begin{table}[h!]
	\centering
	\small
	\begin{tabular}{l|l|l|l|}
		\cline{2-4}
		&  IAM-OnDB &  Ours &  \textbf{Unified}  \\ \hline
		\multicolumn{1}{|l|}{Avg. Age (SD) } & 24.84 ($\pm$ 6.2) & 23.55 ($\pm$ 5.7)  & \textbf{24.85 ($\pm$ 6.19)}  \\ \hline
		\multicolumn{1}{|l|}{Females \%} & 34.00 & 32.63  & \textbf{33.55}  \\ \hline
		\multicolumn{1}{|l|}{Right-handed \%} & 91.50 & 96.84  & \textbf{93.22}  \\ \hline
		\multicolumn{1}{|l|}{\# sentences} & 11242 & 63182  & \textbf{17560}  \\ \hline
		\multicolumn{1}{|l|}{\# unique words} & 11059 & 6418  & \textbf{12718}  \\ \hline
		\multicolumn{1}{|l|}{\# word instances} & 59141 & 26040 &  \textbf{85181} \\ \hline
		\multicolumn{1}{|l|}{\# characters} & 262981 & 143975 & \textbf{406956} \\ \hline
	\end{tabular}
	\caption{Data statistics. In bold, the final \textit{unified} dataset.}
	\label{table:dataset_stats}
\end{table}
\vspace{-0em}

The large number of subjects contained in our dataset allows to capture substantially large variation in styles, crucial to perform any handwriting-related learning task, since handwriting typically exhibits large variation both inter and intra subjects. Furthermore, the dataset contains samples from different digitization mechanisms and should hence be useful in learning models that are robust to the exact type of input.

\begin{figure}[t]
\centering
\includegraphics[width=.9\columnwidth]{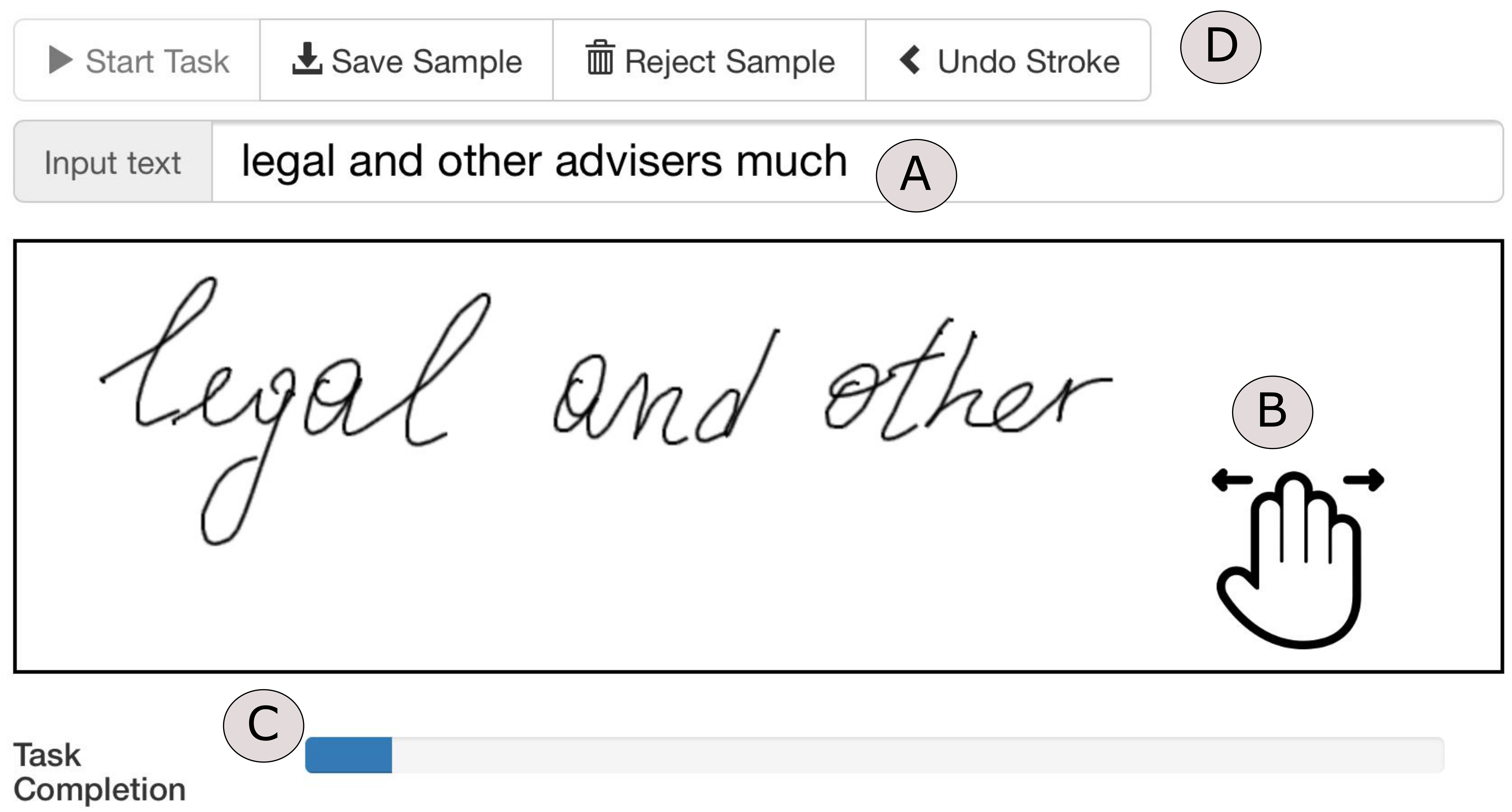}
\caption{Data collection interface. Users are presented with text to write (A) in the scrollable collection canvas (B). A progress bar (C) informs the user on the status. A writer can save, reject or correct samples using the toolbox buttons (D).}
\label{fig:dataset_collection_interface}
\end{figure}

We developed a web tool to collect samples from 94 authors (see \figref{fig:dataset_collection_interface}). Inline with IAM-OnDB we asked each subject to write extracts of the Lancaster-Oslo-Bergen (LOB) text corpus \cite{Stig1986} using an iPad Pro. Besides stylus information, we recorded age, gender, handedness and native language of each author. The data is again segmented at the character level and misspelled or unreadable samples have been removed.

Samples from 200 authors in the dataset stem from the IAM-OnDB dataset and were acquired using a smart whiteboard. 
The data provide stylus information (i.e., x-y coordinates, pen events and timestamps) as well as transcription of the written text, on a per-line basis (\figref{fig:dataset}, A). We purged 21 authors from the original data due to low-quality samples or missing annotations. Furthermore, to improve the granularity of annotations we process the remaining samples, segmenting them down to the character level, obtaining ASCII labels for each character (\figref{fig:dataset}, B and C). For segmentation we used a commercial tool~\cite{MyScript2016} and manually cleaned-up the results.


\section{Conclusion and Future Work}\label{sec:discussion}
We have proposed a novel approach for making digital ink editable, which we call conditional variational recurrent neural networks (\modelname). At the core of our method lies a deep NN architecture that disentangles content from style. The key idea underlying our approach is to treat style and content as two separate latent random variables with distributions learned during training. At sampling time one can then draw from the content and style component to \emph{edit} either style, content or both, or one can \emph{generate} entirely new samples. These learned statistical distributions make traditional approaches to NN training intractable, due to the lack of access to the true distributions. Moreover, to produce realistic samples of digital ink the model needs to perform auxiliary tasks, such as controlling the spacing in between words, character segmentation and recognition. Furthermore, we have build a variety of proof-of-concept applications, including conditional synthesis and editing of digital ink at the word level. Initial user feedback, while preliminary, indicates that users are largely positive about the capability to edit digital ink - in one's own handwriting or in the style of another author.

To enable the community to build on our work we release our implementation as open-source.
Finally, we have contributed a compound dataset, consolidating existing and newly collected handwritten text into a single corpus, annotated to the character level. Data and code are publicly available \footnote{\url{https://ait.ethz.ch/projects/2018/deepwriting/}}.

While our model can create both disconnected and connected (cursive) styles, its performance is currently better for the former, simpler case. This also applies to most state-of-the-art character recognition techniques, and we leave extending our method to fully support cursive script for future work. Further, we are planning to integrate our currently auxiliary character recognition network into the proposed architecture. One interesting direction in this respect would be the inclusion of a full language model. Finally, and in part inspired by initial feedback, we believe that the underlying technology bears a lot of potential for research in other application domains dealing with time-series, such as motion data (e.g., animation, graphics) or sketches and drawings (e.g., arts and education).


\section{Acknowledgements}\label{sec:ack}
This work was supported in parts by the ERC grant OPTINT
(StG-2016-717054). We thank all the participants for their time and efforts in taking part in our experiments.

\balance{}

\bibliographystyle{SIGCHI-Reference-Format}
\bibliography{deep-writing-chi18}

\section{Appendix}
\label{sec:appendix}

\subsection{Alphabet}
We use the following numbers, letters and punctuation symbols in our alphabet: \\
$0123456789abcdefghijklmnopqrstuvwxyzABCDEFGHIJK \\ LMNOPQRSTUVWXYZ'.,-()/$
\subsection{Data Preprocessing}
\begin{enumerate}
\item In order to speed up training we split handwriting samples that have more than $300$ strokes into shorter sequences by using $eoc$ labels so that the stroke sequences of letters remain undivided. We create $705$ validation and $34577$ training samples with average sequence length 261.012 ($\pm$ 85.1).
\item Pixel coordinates ($u_0$, $v_0$) of the first stroke is subtracted from the sequence such that each sample starts at the origin $(0,0)$. 
\item By subtracting $x_{t}$ from $x_{t+1}$, we calculate the changes in 2D-coordinate space and use relative values in training. Note that the $pen-up$ events remained intact.
\item Finally, we apply zero-mean unit-variance normalization on 2D-coordinates by calculating \textit{mean} and \textit{std} statistics on the whole training data.
\end{enumerate}

\subsection{Network Configuration}
Both $\tau^{inp}$ and $\tau^{latent}$ are LSTM cells with $512$ units. Similarly, feed-forward networks $g^*$ consists of $1$-layer with $512$ units and \textit{ReLu} activation function. We use $32$-dimensional isotropic Gaussian distributions for latent variables $z^p$, $z^q$ and for each GMM component. 

Our BiRNN classifier consists of $3$-layer bidirectional LSTM cells with $512$ units. A $1$-layer fully connected network with $256$ units and \textit{ReLu} activation function takes BiRNN representations and outputs class probabilities. 

\subsection{Training}
We use ADAM optimizer with default parameters. Learning rate is initialized with $0.001$ and decayed exponentially with a rate of $0.96$ after every $1000$ mini-batches. We train our model for $200$ epochs by using a mini-batch size of $64$.

\subsection{Additional Results}
\begin{figure}[t!]
	\centering
	\includegraphics[width=0.99\linewidth,trim={5pt 250pt 0pt 0pt},clip]{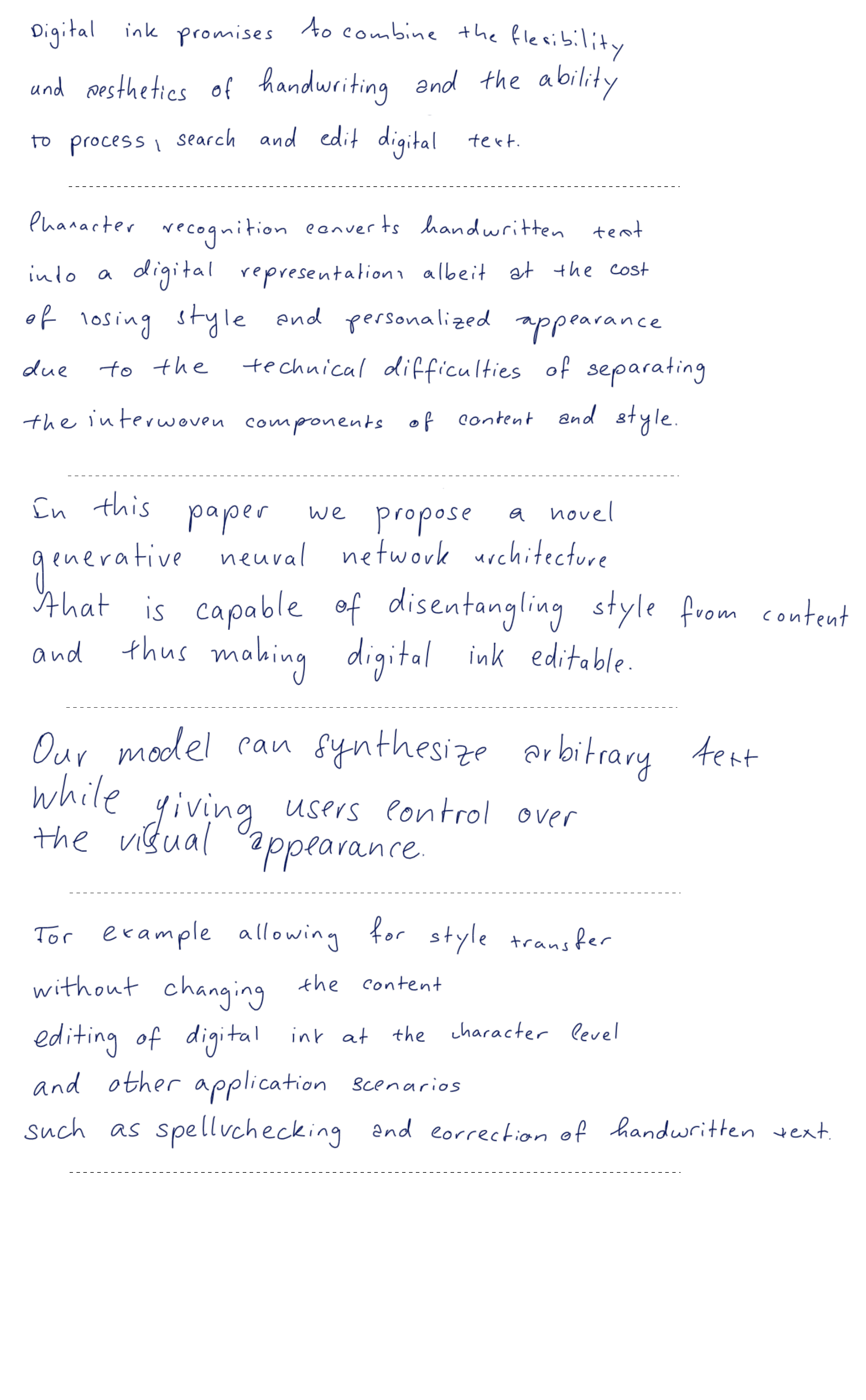}
	\caption{Handwritten text synthesized from the abstract of this paper. Each sentence is ``written'' in the style of a different author.}
	\label{fig:interface-synthesis-full}
\end{figure}

As an additional result, we have synthesized the entire abstract from this paper using a different style per sentence. The result is shown in \figref{fig:interface-synthesis-full}, and illustrates how our model is able to generate a large variety of styles.

\end{document}